\documentclass[conference]{IEEEtran}
\IEEEoverridecommandlockouts
\usepackage{cite}
\usepackage{amsmath,amssymb,amsfonts}
\usepackage{algorithmic}
\usepackage{graphicx}
\usepackage{textcomp}
\usepackage{hyperref}
\usepackage{xcolor}
\usepackage{comment}
\usepackage{xspace}
\usepackage{enumitem}
\usepackage{caption,subcaption}
\usepackage{siunitx}
\usepackage{booktabs}

\def\BibTeX{{\rm B\kern-.05em{\sc i\kern-.025em b}\kern-.08em
    T\kern-.1667em\lower.7ex\hbox{E}\kern-.125emX}}

\newcommand*\Reactivatenumber{%
  \lst@AddToHook{OnNewLine}{%
   \let\thelstnumber\origthelstnumber%
   \advance\c@lstnumber\@ne\relax}%
}

\newcommand{\ffis}{FFIS\xspace}
\newcommand{\ffisnospace}{FFIS}

\begin{document}

\title{Characterizing Impacts of Storage Faults on HPC Applications: A Methodology and Insights}


\author{
  \IEEEauthorblockN{Bo Fang,\IEEEauthorrefmark{1}\textsuperscript{\IEEEauthorrefmark{3}}
  Daoce Wang,\IEEEauthorrefmark{2}\textsuperscript{\IEEEauthorrefmark{3}}
  Sian Jin,\IEEEauthorrefmark{2}
  Quincey Koziol,\IEEEauthorrefmark{4}
  Zhao Zhang,\IEEEauthorrefmark{5}\\
  Qiang Guan,\IEEEauthorrefmark{6}
  Suren Byna,\IEEEauthorrefmark{4}
  Sriram Krishnamoorthy,\IEEEauthorrefmark{1}\IEEEauthorrefmark{2}
  Dingwen Tao\IEEEauthorrefmark{2}\textsuperscript{\IEEEauthorrefmark{7}}}
  \IEEEauthorblockA{\IEEEauthorrefmark{1}
  Pacific Northwest National Laboratory,
  Richland, WA, USA}
  \IEEEauthorblockA{\IEEEauthorrefmark{2}
  Washington State University,
  Pullman, WA, USA}
  \IEEEauthorblockA{\IEEEauthorrefmark{4}
  Lawrence Berkeley National Laboratory,
  Berkeley, CA, USA}
  \IEEEauthorblockA{\IEEEauthorrefmark{5}
  Texas Advanced Computing Center,
  Austin, TX, USA}
  \IEEEauthorblockA{\IEEEauthorrefmark{6}
  Kent State University,
  Kent, OH, USA}
 }



\maketitle

\begingroup\renewcommand\thefootnote{\IEEEauthorrefmark{3}}
\footnotetext{These authors contributed equally.}
\endgroup

\begingroup\renewcommand\thefootnote{\IEEEauthorrefmark{7}}
\footnotetext{Corresponding author: Dingwen Tao (\href{mailto:dingwen.tao@wsu.edu}{dingwen.tao@wsu.edu}).}
\endgroup

\pagestyle{plain}

\setlength{\textfloatsep}{3pt}

\begin{abstract}
In recent years, the increasing complexity in scientific simulations and emerging demands for training heavy artificial intelligence models require massive and fast data accesses, which urges high-performance computing (HPC) platforms to equip with more advanced storage infrastructures such as solid-state disks (SSDs). While SSDs offer high-performance I/O, the reliability challenges faced by the HPC applications under the SSD-related failures remains unclear, in particular for failures resulting in data corruptions. The goal of this paper is to understand the impact of SSD-related faults on the behaviors of complex HPC applications. To this end, we propose \ffis, a FUSE-based fault injection framework that systematically introduces storage faults into the application layer to model the errors originated from SSDs. \ffis is able to plant different I/O related faults into the data returned from underlying file systems, which enables the investigation on the error resilience characteristics of the scientific file format. We demonstrate the use of \ffis with three representative real HPC applications, showing how each application reacts to the data corruptions, and provide insights on the error resilience of the widely adopted HDF5 file format for the HPC applications.
\end{abstract}

\section{Introduction}
\label{sec:intro}
In recent years, the increasing complexity in scientific simulation and emerging demands for training heavy artificial intelligence (AI) models require massive and fast data accesses, which urges high-performance computing platforms to equip more advanced storage infrastructures. To this end, flash-based solid-state drives (SSDs) have been widely employed in HPC systems as a replacement of hard disk drives (HDDs) to achieve an order of magnitude speedup in data access. Prior studies~\cite{flash-field,ssdfailures,bianca-flash} showed that this rapid adoption of SSD-based I/O components, however, raises a new challenge to the overall HPC reliability. As shown in~\cite{flash-field}, uncorrectable bit error rate (UBER) of data center SSDs are between \num{e-11}, \num{e-9}, which 
results in
a collective high error rate on the large-scale HPC system and breaks the JEDEC 2016 requirement for enterprise class ( <\num{e-16})~\cite{jedec}. Such concern is expected to be continuously present due to the fact that the sources of SSD failures, i.e., cell wear-out, program/read disturb errors, retention errors, power faults, radiations, etc, would not dissolve shortly.

There are two classes of failures prominently experienced on SSDs: fail-stop and partial disk failures. Unlike the fail-stop failures that cause the SSDs to become inaccessible from the higher level of the stack, partial drive failures only affect a portion of the SSD operations and the device remains to work from the user's perspective~\cite{partial-1,partial-2,partial-3,partial-4,partial-5,partial-6,partial-7,partial-8}. The consequence of the partial failures is severe: they may cause corrupted data on SSDs and trigger issues in the file system and application layers in the I/O stack. Therefore, HPC applications need to mitigate the impact of the partial disk failures and tolerate the data corruption propagation due to such failures from SSDs.    

Mitigating data corruption in an HPC application has been a challenging task that requires efficient and effective solutions. Towards this goal, a wide spectrum of studies~\cite{Ashraf2015,flipit} employ statistical fault injections to characterize the impact of hardware faults, and take the characteristics obtained on per application basis to guide the design of the fault tolerance strategy. However, since the common fault models investigated in those studies are bit-flip faults affecting computational units and memory, their insights are not indicative to reason about the application's error resilience characteristics against SSD-related data corruption, which demands a new systematic characterization approach.   

The goal of this paper is to provide methodology for characterizing how different types of SSD-related faults would affect the behaviors of the HPC applications. We mimic the impact of partial disk failures on applications by introducing faults via an application's I/O operations and observe the outcomes of the applications after the fault injection. We introduce a fault injection framework, \ffis \footnote{https://github.com/FabioGrosso/pFsysFI.} (\underline{F}USE-based \underline{F}ault \underline{I}njection for \underline{S}torage) that leverages the FUSE~\cite{fuse} interface to systematically inject faults into the applications' I/O path. \ffis supports multiple fault models, each of which represents a specific data corruption scenario observed from the partial disk failures. \ffis offers a uniform interface for users to apply fault injection campaigns on various real HPC applications, without any modification on the applications.


\ffis is built based on the following key assumptions: \textit{(i)} we focus on the data corruption that manifests on the application level, such as bit flips, shorn writes, etc. {We explain the details of these faults in Section~\ref{sec:method}}.\textit{(ii)} those errors can further propagate beyond the file system layer~\cite{eval-ssd,pfault,errfs}, skip the verification mechanisms (e.g., fsck~\cite{fsck}) and silently compromise the data integrity of the applications. 

Enabled by \ffis, systematic fault injection studies can be 
performed 
to reveal error resilience characteristics of an application or the common components exercised across applications: \textit{(i)} \ffis is able to evaluate different inherent error masking capabilities for different applications, or to measure such ability of each phase of an application.
This suggests a potential trade-off space for HPC systems to lower the requirement of the SSDs' reliability for faster data access while maintaining the same level of the overall application's reliability;  \textit{(ii)} as the modern HPC applications tend to leverage the scientific file format for efficient data management, \ffis is able to investigate how the certain scientific file format library handles the storage errors affecting both the file metadata and application data, thus to obtain the possible common error resilience characteristics of the applications while operating on such file format. This paper makes the following contributions:
\begin{itemize}
    \item We design and build a fault injection framework, called \ffis, 
    to model SSD-related failures at software level and to inject such faults systematically into HPC applications.
    
    \item We apply \ffis on three real-world HPC applications through comprehensive, large-scale fault injection experiments. Our evaluation shows that applications exhibit distinct error resilience characteristics for different fault models, and we offer the detailed explanation for each application's unique resilience characteristics.    
    
    \item We unveil application-specific behaviors operating on the most widely adopted scientific file format - HDF5, and show the fault-tolerance
    behaviors of the HDF5 library against errors affecting the HDF5 metadata. We identify certain fields in the metadata that may cause SDC outcomes when affected by faults, and provide the solutions for auto-correction. To the best of our knowledge, it is the first research effort to systematically characterize the detailed error resilience of the HDF5 file format.   
\end{itemize}


\section{Background}
\label{sec:background}
In this section, we describe the key context for our study and 
its importance
to 
motivate the FFIS framework development.


\textit{Fault, error, and failure}: As described in~\cite{terms}, a (hardware) fault, error, and failure chain is defined as the following events: ``a service is a sequence of a system’s external states, a service \textbf{failure} means that at least one (or more) external state of the system deviates from the correct service state. The deviation is called an \textbf{error}. The adjudged
cause of an error is called a \textbf{fault}''. Below we specify these terms in the different context:

\begin{itemize}
    \item \textit{Storage system}: an SSD's partial failures refer to the events where SSDs are not providing the expected behavior or outcomes such that the SSDs' internal states are left with flipped bits or shorn data, and what causes such failures are considered as hardware faults, including power faults or ratification faults, etc.. 
    \item \textit{File system}: the file system failures refer to the unsuccessful file operations such as I/O errors returned to the application, which can result from the storage failures, or software bugs (not covered in this study). 
    \item \textit{Application}: a failure of an application refers to that scenario that the outcome of the application differs from the expected: the application either terminates before it finishes (i.e., \textit{crash}), or it suffers from data corruption. If the application is able to identify the errors, this failure is categorized as \textit{detected}, otherwise such data corruption becomes silent data corruption (SDC). 

\end{itemize}

\textbf{\textit{FUSE}} File system in user-space is the most widely used user-space file system framework~\cite{fuse} on Unix/Linux systems. It exposes the file operations to the file system users and allows the users to implement their own file operations such as \textit{open(), read(), write(),etc.} if needed. When a FUSE file system is in use, the programs are able to access the data using the standard file operation system calls (i.e., POSIX) through the implementation of those system calls in the FUSE. 

\textbf{\textit{Why choose FUSE}} We choose FUSE as the underline file system interface for the fault injection framework for two reasons: \textit{(i)}, as the goal is to study the impact of the failures on applications, the fault injection framework focuses on mimicking the SSD-related faults occurring during the I/O operations in the application level. FUSE allows us to implement such faults with a relatively straightforward manner;\textit{(ii)}, since a FUSE-based file system works as the callbacks for the file operations, it allows the applications to call the user-implemented I/O primitives without modification on either the application or the actual running file system. This releases the burden for HPC applications that usually consists of complex behaviors and conservative execution environment.


\textbf{\textit{Manifestation of SSD failures}} Zheng et al.~\cite{powerfault} found that the power faults can cause a large number of SSDs to fail partially and produce the number of chip-level bit errors that exceed the correction capability of SSDs' error correcting code (ECC) and bypass the SMART (Self-Monitoring, Analysis and Reporting Technology) system~\cite{smart}. For example, A recent study~\cite{flash-field} reports that on SSDs the occurrence of partial drive failures that lead to data corruption can be an order of magnitude higher than on HDDs~\cite{flash-field}. As shown in~\cite{eval-ssd}, this type of the SSD failures can manifest in the file system and affect the application's I/O behaviors. This leads to two classes of failures in general: \textit{(a)} the file system throws the I/O errors and leaves the handling to the application and the typical failures include uncorrectable bit corruption, metadata corruption, incomplete I/O operations; \textit{(b)} the file system does not detect such failures and the failures may cause data corruption in the application. These failures include silent bit corruption (bit flips in the data), shorn write (a write operation is partially done on the device), and dropped write (the file system issues the write but never gets executed on the device), which is the focus of this paper. 

\textbf{\textit{Why focusing on HDF5 file format}} 
Hierarchical Data Format version 5 (HDF5) provides an API for performing I/O, data management tools, and a portable file format \cite{hdf5}. In \cite{Byna:ExaHDF5:2020}, the authors show that HDF5 is the most used I/O library on HPC systems at the National Energy Research Scientific Computing Center (NERSC) and at several US Department of Energy(DOE) supercomputing facilities, including the Leadership Computing Facilities (LCFs). HDF5 has a rich ecosystem and various third-party bindings are available to manage data \cite{hdf5-wiki}. For instance, Matlab uses HDF5 as the primary storage format \cite{hdf5-wiki} to allow the operations on the numeric data. Considering this rich eco-system and popular usage, verifying the resilience of the file format with \ffis offers significant insights for both HDF5 and application developers.

\begin{figure}
    \centering
    \includegraphics[width=0.45\textwidth]{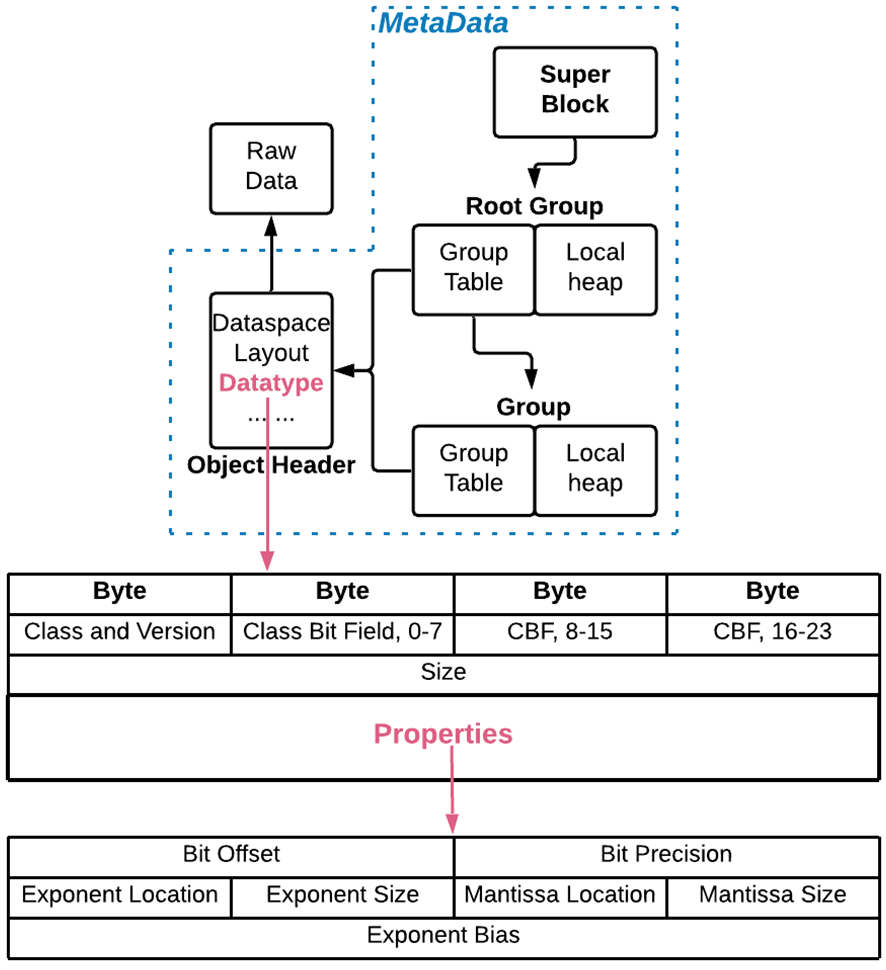}
    \caption{Overview of HDF5 file structure (top), layout of datatype message (middle), floating-point property (bottom).}
    \label{fig:h5-struc}
\end{figure}

\textbf{\textit{HDF5 file structure}} 
Figure~\ref{fig:h5-struc} depicts a general structure of the HDF5 format~\cite{hdf5-ff}.
The HDF5 format defines a cascading style metadata: it holds a super block that points to multiple groups, and each group represents an object header that may store other groups or datasets within it. A dataset represents the actual data contained within the HDF5 file. It contains an object header that describes the data space, the data type, the data layout, and other useful information. Internally, HDF5 uses B-tree nodes to index where a particular information is stored. For example, we show the detailed layout and relation of datatype message (middle), and floating-point property (lower) in Figure~\ref{fig:h5-struc}.

\section{\ffis framework}
\label{sec:fifaa}
In this section, we present the design and implementation of our fault injection framework - 
\ffis. We first describe the overall system design of \ffis with the focus on the general workflow interacting with the FUSE file system I/O path. Then, we describe the fault models currently \ffis supports, and finally, we explain each component of \ffis and highlight the key features for systematically conducting error resilience characterization on HPC applications.

\subsection{Design Goal}

The goal of \ffis is to mimic the SSD failures as 
software-implemented faults that corrupt the application data in a controlled manner. To this end, there are four requirements that \ffis needs to follow:
\begin{itemize}
    \item {\it Transparency:} \ffis should be able to transparently plant a fault into an application at runtime without modifying the application code (R1). This requires that \ffis should not make any assumptions about what specific file systems that the application work with or what specific I/O operations the application needs, and ensures that the application's calls to the I/O system remains unaffected due to any fault injection framework's artifacts.   
    
    \item {\it Convenience:} As HPC applications tend to be conservative and sensitive to the computation environment, the \ffis framework should be deployed without requiring modification the compilation/execution environment of the application (R2).   
    
    \item {\it Comprehensiveness:} \ffis should support multiple fault models that represent different types of SSD-related failures. This extends the bit-flip based fault models (R3). 
    
    \item {\it Repressiveness:} as the goal of \ffis is to mimic the SSD-related failures at software level, it is important that \ffis can introduce the faults uniformly over the set of all corresponding file operations (R4). 

\end{itemize}

\begin{figure}[h]
    \centering
    \includegraphics[width=0.5\textwidth]{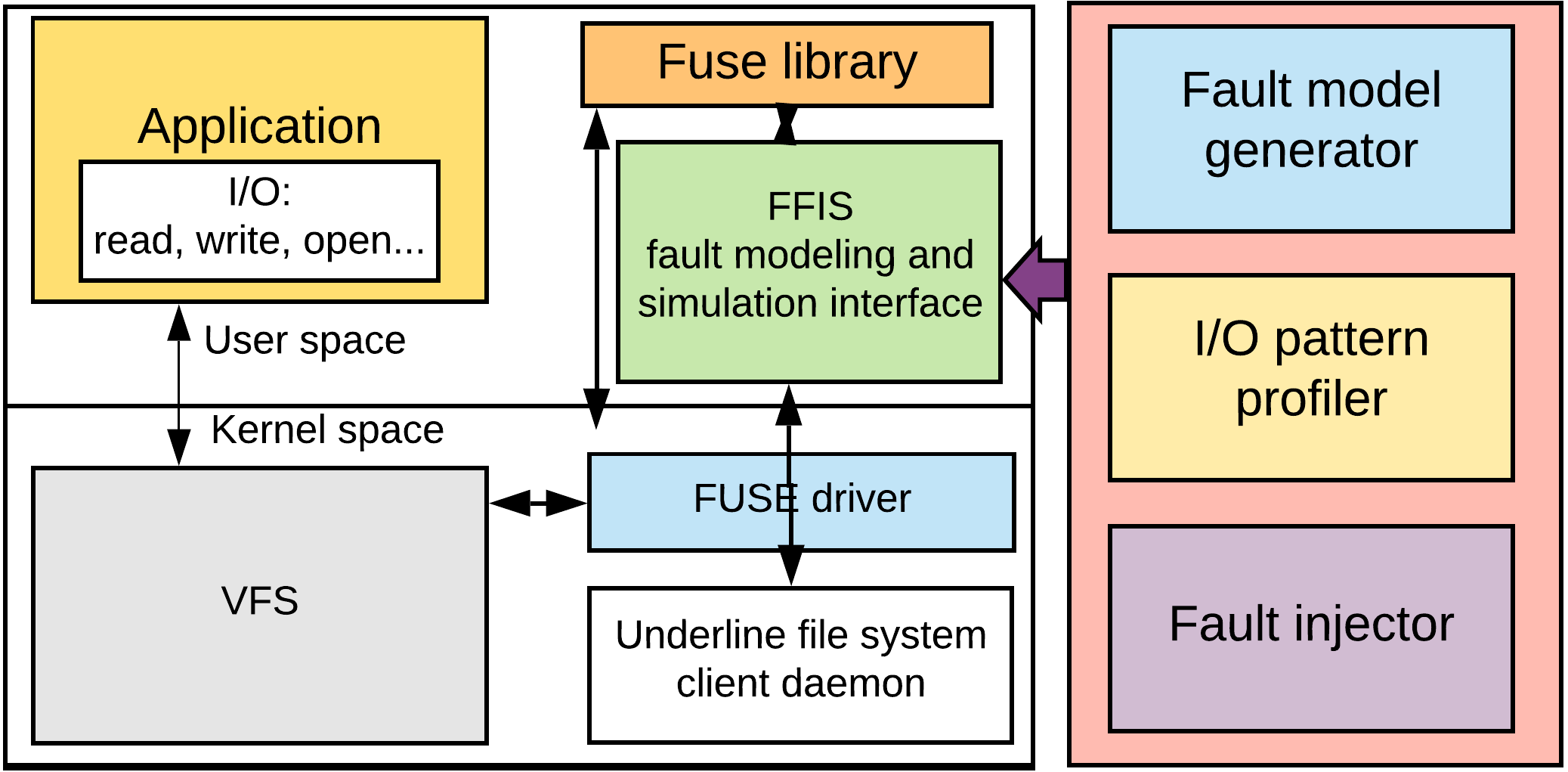}
    \caption{Overview of \ffis framework based on FUSE.}
    \label{fig:fifaa}
\end{figure}

In Figure~\ref{fig:fifaa}, we illustrate the overview design of the \ffis framework. \ffis employs the FUSE interface to introduce a standard file system for the application to use. The application under the characterization runs on top of the \textit{FFISFS} file system, which can work in parallel with the actual file systems exercised by the application such as lustre, GPFS, ext4, etc. \textit{FFISFS} works similarly to what normal FUSE-based file system does: at the time the \textit{FFISFS} file system is mounted, the file system handler is registered with the OS kernel. If an application issues, for example read/write/stat requests for the mounted \textit{FFISFS}, the kernel forwards these IO-requests to the handler and then sends the handler's response back to the user. 

That said, \ffis intercepts the I/O system calls via instrumenting the file system primitives inside the FUSE interface without any change on the application code, which addresses the R1. Since FUSE offers the uniform API that is agnostic to how the application invokes the I/O, \ffis is able to support different applications without worrying about the specific set of APIs that the application employs. To invoke the \ffis framework for fault injection, the application only needs to make sure that the requested file(s) reside in the mount point of \textit{FFISFS}, which addresses the R2. 

\subsection{Fault Model}
\label{sec:fm}

\ffis currently supports three types of faults, each of which corresponds to a SSD failures` manifestation: \textsc{bit flip}, \textsc{shorn write}, and \textsc{dropped write}. As discussed in Section~\ref{sec:background}, these fault models could cause different types of data corruption in applications. Table~\ref{tab:faultmodel} summarizes what file system operations can potentially be used to host the fault for each fault model and the implementation specification \ffis defines for such fault. These specifications are aligned with the observations from the prior studies~\cite{powerfault,eval-ssd}.

\begin{table}[th]
\caption{Fault models supported by \ffis. \small{The table reports the examples of the FUSE primitives where the faults can manifest, and the key features \ffis implements for each fault model.}}
\resizebox{\columnwidth}{!}{ 
\begin{tabular}{|l|l|l|}
\hline
\textbf{Fault model} & \textbf{Examples of Affected FUSE primitives} & \textbf{Features} \\ \hline
Bitflip & \textit{\ffisnospace\_write, \ffisnospace\_mknod,\ffisnospace\_chmod ...} & flip consecutive multiple bits\footnote{Prior study~\cite{}  shows that the most common multi-bit faults experienced in memory occur in consecutive bits.} \\ \hline
Shorn write & \textit{\ffisnospace\_write, \ffisnospace\_mknod, \ffisnospace\_chmod ...} & \begin{tabular}[c]{@{}l@{}}completely write the first 3/8th of 4KB \\ block or first 7/8th of 4KB block to \\ the device at the granularity of 512B\end{tabular} \\ \hline
Dropped write & \textit{\ffisnospace\_write, \ffisnospace\_mknod, \ffisnospace\_chmod ...} & the write operation is ignored \\ \hline
\end{tabular}
}
\label{tab:faultmodel}
\end{table}

\begin{figure}
     \centering
     \begin{subfigure}[b]{0.49\linewidth}
         \centering
         \includegraphics[width=\linewidth]{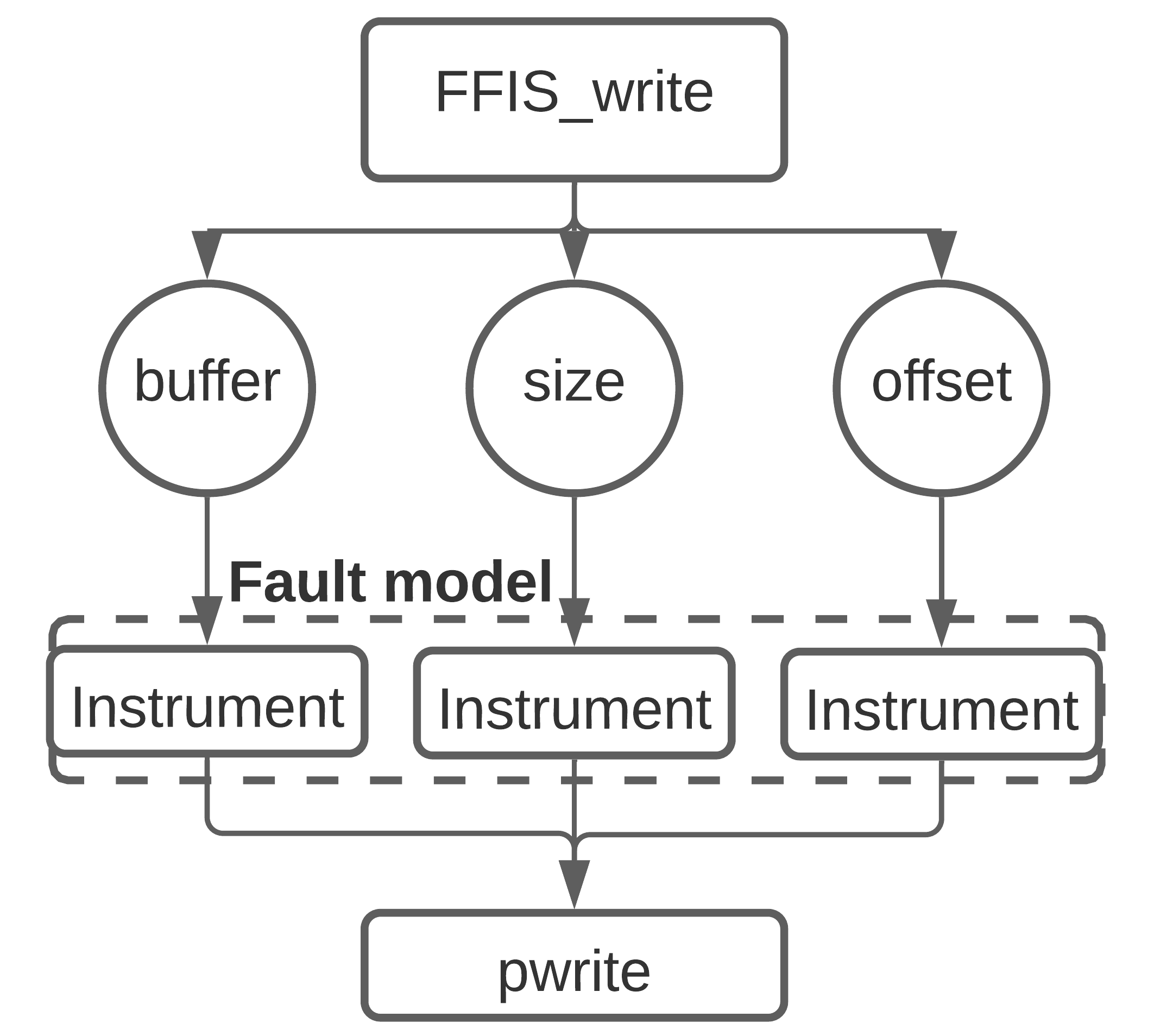}
         \caption{The FFIS instrumentation on \textit{FFIS\_write} primitive. It modifies the content of the \textsc{buffer, size and offset} passed to the \textit{FFIS\_write}, and the modified content is fed to the system call \textit{pwrite}.}
         \label{fig:fwrite}
     \end{subfigure}
     \hfill
     \begin{subfigure}[b]{0.45\linewidth}
         \centering
         \includegraphics[width=\linewidth]{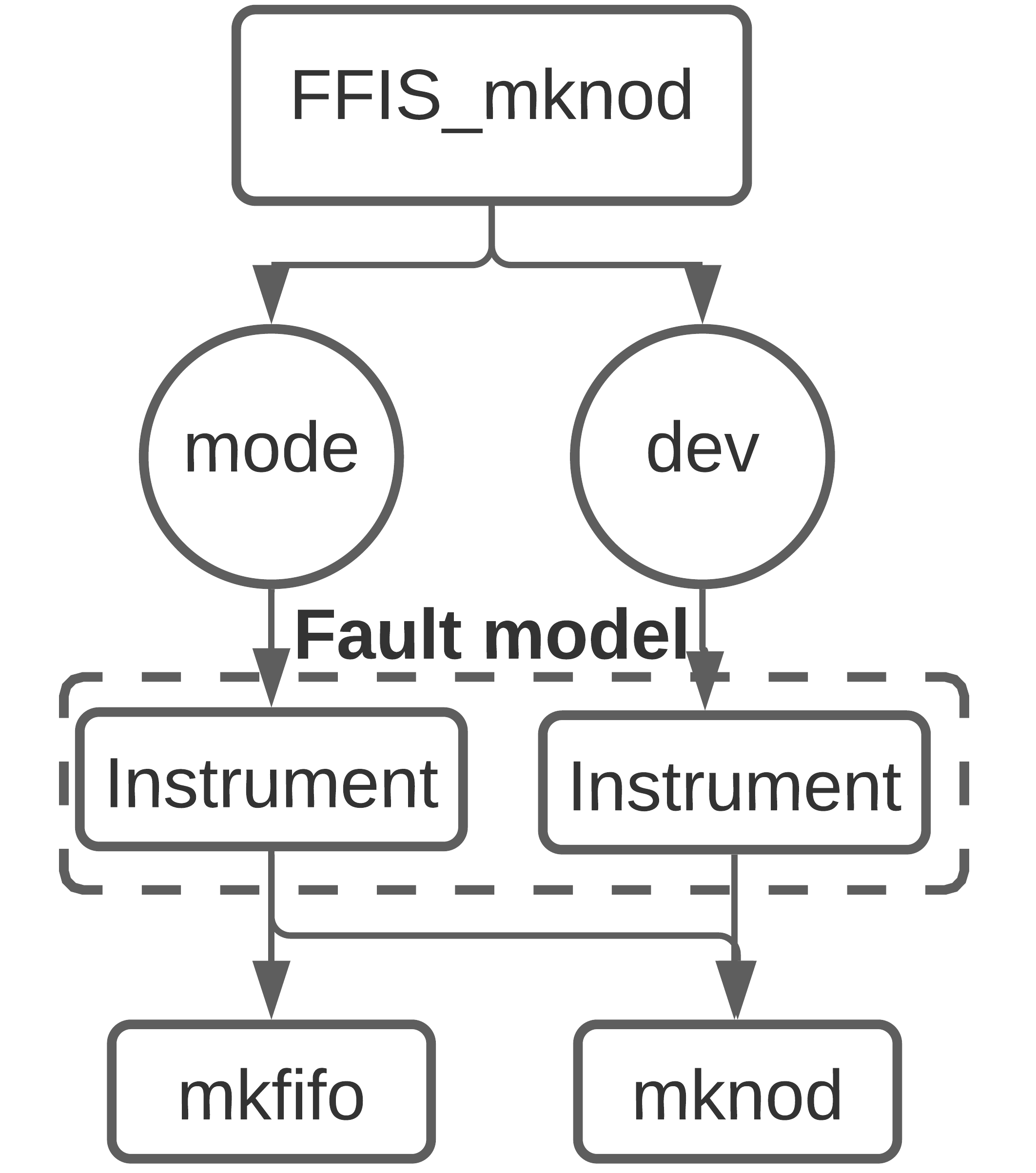}
         \caption{The FFIS instrumentation on \textit{FFIS\_mknod} primitive. It modifies the content of the \textsc{mode, and dev} passed to the \textit{FFIS\_write}, and the modified content are fed to the system call \textit{mknod} and \textit{mkfifo}.}
         \label{fig:fmknod}
     \end{subfigure}
        \caption{Examples of how \ffis performs fault injection by modifying content of target primitives based on fault models.}
        \label{fig:impl}
\end{figure}

\ffis instruments the primitives and modifies the file state either in the metadata structures or the data blocks. Figure~\ref{fig:impl} shows two examples of the instrumentation strategy performed by \ffis. In particular, for the \textit{FFIS\_write} primitive, \ffis modifies the parameters of the \textit{FFIS\_write} depending on the fault model and pass the modified content to the underline file system interface - \textit{pwrite} in this case (In Section~\ref{sec:method} we explained the details of the instrumentation on \textit{FFIS\_write}); For the \textit{FFIS\_mknod} primitive, it conducts the similar operations on the parameters based on the fault model, and the modified data are sent to \textit{mknod} and \textit{mkfifo} system calls respectively.

\subsection{\ffis Workflow}
\begin{figure}[h]
    \centering
    \includegraphics[width=0.42\textwidth]{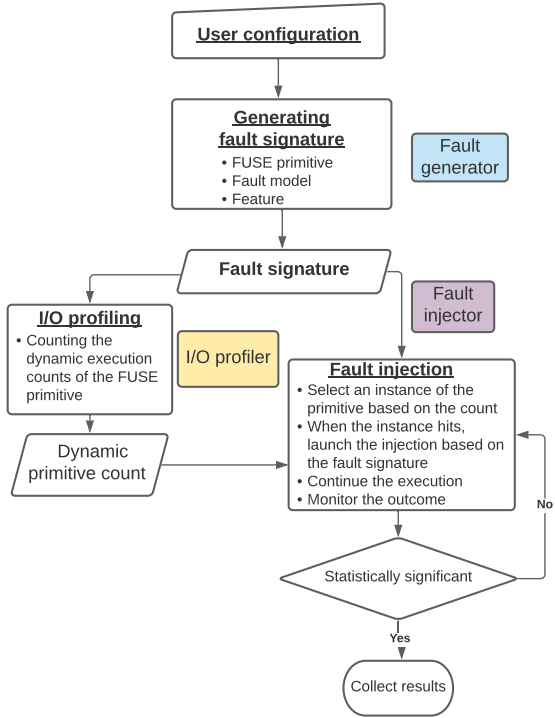}
    \caption{\ffis workflow that illustrates the process of systematically introducing a fault into the application's I/O path.}
    \label{fig:workflow}
\end{figure}

\ffis consists of three components: the fault generator, the I/O profiler and the fault injector. Below we explain details of each component, and depict the overall workflow of the \ffis framework in Figure~\ref{fig:workflow}.

\textbf{Fault Generator} The fault generator reads the configuration specified by the user to produce a fault signature, which includes the fault model, the file system primitive where the fault would be injected for that fault model, and the choice of the feature associated with the fault model. The fault signature then gets passed to both the I/O profiler and the fault injector.

\textbf{I/O Profiler} The goal of the I/O profiler is to count the number of times that the primitive (i.e. configured in the fault signature) gets executed during the execution. To this end, the I/O profiler instruments the primitive inside the FUSE and executes the application fault-free to obtain the total count. It then passes this dynamic count of the primitive to the fault injector.  

\textbf{Fault Injector} The fault injector performs the actual fault injection operations with the fault signature, including the  fault model, the feature and the primitive, and the dynamic count obtained from the profiler. For each fault injection run, it first generates a random number from 0 to count-1, and executes the application normally. When the execution count of the target primitive hits that random number, the fault injector applies the fault based on the fault signature. This process is repeated until the statistical significance is reached. 

In summary, the whole workflow of the fault injection proceeds as follows: \ffis loads the user configuration of the application, the fault model, the target primitive and the feature of the fault model, and launches the application in the mount point of FFISFS for profiling. Once it obtains the total executed count for the target primitive, it enters the fault injection mode: for each fault injection run, it randomly chooses an instance of the execution of the primitive, and plant the fault based on the fault signature with the fault specification described in Section~\ref{sec:fm}. Note that in each run, FFISFS would be mounted and unmounted to mimic the real scenario on the HPC system for the application. 

\section{Evaluation Methodology}
In this section, we present our evaluation methodologies.
We first introduce our experimental platform and then present our fault models and their specifications. Lastly, we describe our evaluation applications and our fault injection approaches.

\label{sec:method}
\subsection{Experimental Platform}
We perform our fault injection experiments using a local server with 24-core AMD Ryzen Threadripper 3960X CPUs. We also use the Frontera system \cite{frontera} (equipped with two Intel Xeon Platinum 8280 CPUs each node) at the Texas Advanced Computing Center for simulations and post-analyses. 

\subsection{Fault Model Specification}

As discussed in Section~\ref{sec:fifaa}, the fault injection experiments incorporate a fault signature to determine the fault model based on the FUSE primitive and the fault feature. In this study, we consider three fault models:
\textsc{bit flip}, \textsc{shorn write}, and \textsc{dropped write}. To efficiently and directly study the impact of data corruption, we choose the \textit{\ffisnospace\_write} primitive to implement all the fault models, while each fault model has its own feature and the implementation strategy:

\begin{itemize}
    \item \textsc{bit flip}: \ffis flips 2 consecutive bits randomly chosen in the buffer that is used in \textit{pwrite} system call inside the \textit{\ffisnospace\_write} for each fault injection run.\footnote{All the bit-flip results shown in the paper are based on the observation associated with this fault model. We also tested the 4-bit bit flip model and the SDC rate remains minimal for Nyx.}  
     \item \textsc{shorn write}: \ffis modifies a write operation to lose the last $\frac{1}{8}$th of the data by stripping down the buffer used in \textit{pwrite} system call. While the buffer shrinks, the size is yet kept as the original value: this would introduce undefined data to write to the file system, which copes with the shorn write failure on SSDs.
     \item \textsc{dropped write}: \ffis ignores the \textit{pwrite} call for that instance inside the \textit{FFIS\_write} and sets the return value of the \textit{FFIS\_write} to the original size of the data buffer.   
\end{itemize}

\subsection{Evaluation Applications and Fault Injection Approaches}
We conduct our experiments on three representative real-world HPC applications: Nyx \cite{almgren2013nyx}, QMCPACK \cite{kim2018qmcpack}, and Montage \cite{jacob2009montage}. The common feature of these three applications are that they all have a large number of I/O operations and their own post-analysis processes. A large number of I/O operations mean that there could be more I/O faults, and the post-analysis process defines the impacts of these faults on each application. We use Table \ref{tab:benchmark} to list some key information of our evaluation benchmark applications. For each application's dataset under each fault model, we conduct 1,000 fault injection runs to obtain a statistically significant estimate, which leaves a 1\%$\sim$2\% error bar on average for 95\% confidence interval.


\begin{table*}[ht]
\caption{Description of tested HPC applications.}
\begin{tabular}{|l|l|l|l|l|}
\hline
\textbf{Benchmark} & \textbf{Domain} & \textbf{Package Size} & \textbf{LoC} & \textbf{Method} \\ \hline
Nyx & Astrophysics & 71.9MB & 21K & Adaptive mesh refinement (AMR) based cosmological simulation \\ \hline
QMCPACK & Quantum Chemistry & 381MB & 403K  & Quantum Monte Carlo simulation for electronic structures of molecules \\ \hline
Montage & Astronomy & 126MB & 31K  & Astronomical image mosaic\\ \hline
\end{tabular}
\label{tab:benchmark}
\end{table*}

\subsubsection{Nyx}
Nyx \cite{almgren2013nyx} is an adaptive mesh, hydrodynamics algorithm designed to model astrophysical reacting flows. This code models dark matter as discrete particles moving under the influence of gravity. The fluid in gas-dynamics is modeled using a finite-volume methodology on an adaptive set of 3-D Eulerian mesh. The mesh structure is used to evolve both the fluid quantities and the particles via a particle-mesh method.

It is worth mentioning that Nyx has multiple post-analysis programs, such as power spectrum (statistically describing the amount of the Universe at each physical scale) and dark matter halos (over-densities in the dark matter distribution). In this work, we select the most popular post-analysis: \textsc{halo finder} (based on the Friends-of-Friends algorithm \cite{davis1985evolution}), which aims to find the halos using the ``baryon density'' field in the dataset and output the positions, the number of cells, and mass for each halo it finds, respectively. 

\textbf{Outcome Classification}
Similarly, we compare the output of the halo finder (e.g., NVB\_integral\_512 for $512\times512\times512$ Nyx datasets) of the fault injected case with the original output. If they are bit-wise identical, they are classified as benign. If they differ, and there is no halo found, the cases are detected and otherwise they are the SDC. 

\subsubsection{QMCPACK}
QMCPACK \cite{kim2018qmcpack} is an open-source, high-performance electronic structure application that implements numerous Quantum Monte Carlo (QMC) algorithms for electronic structure calculations of molecular, periodic 2D/3D solid-state systems. Real-space Variational Monte Carlo (VMC), Diffusion Monte Carlo (DMC), and other advanced QMC algorithms are implemented in this package.

\textbf{Fault Injection Preprocess}
We choose the Diffusion Monte Carlo (DMC) algorithm in QMCPACK. It first runs VMC to generate a set of walkers and then performs DMC. Finally, there will be two types of output files -- 000 for the VMC algorithm and 001 for the DMC algorithm -- both of which contain large numbers of floating-point data.

We then use the \textsc{qmca} tool in QMCPACK to obtain the total energies and related quantities. Since there is only one electron of each spin, DMC is supposed to reproduce the exact non-relativistic ground state energy (-2.90372 Hartree) \cite{qmc-he}. 
Thus, by analyzing the energy change after the fault injection, we can observe the impacts of the faults on QMCPACK.

\textbf{Outcome Classification}
To define the class of an outcome, we first bit-wise compare the output file He.s001.scalar.dat of each fault injected case with the original (fault-free) output file. If the two files are the same, this case will be identified as benign. Next, if this case is not benign, we continue to perform QMCPACK's post-analysis to obtain an energy value. Since QMCPACK itself allows minor confidence intervals, we rely on such information to define the boundary between SDC and detected. After communicating with the QMCPACK development team, we decide that if the final energy value remains in [-2.91, -2.90], we will define this case as SDC; otherwise as detected.

\subsubsection{Montage}
\label{mon:spe}
Montage \cite{jacob2009montage} is a toolkit to assemble Flexible Image Transport System (FITS) images into custom mosaics. In this paper, we create a mosaic of 10 2MASS Atlas images in a 0.2-degree area around m101 in the J band, reprojecting them into the TAN projection. We generate both background-matched and uncorrected versions of the mosaic. The final product includes two mosaics of m101 and their corresponding area images, which are used by Montage when co-adding images together to form a mosaic. Specifically, it takes Montage ten stages to generate the final image from the beginning, four in which involves a large amount of I/O reads and writes. So, we inject the faults in different stages and then analyze the final results to study the fault propagation in this program. 


\textbf{Outcome Classification}
To determine each category of the outcomes, we first bit-wise compare the output image m101\_mosaic.jpg of each fault injected case with the original (fault-free) output image. If the two images are the same, the case is identified as benign. 
Then, to determine if a faulty case is SDC or detected, we analyze the output of the process that uses the \textit{fits} file to generate image in the last step. As we (1) observe that the ``min'' value in the output greatly correlates with the correctness of the final image, and (2) consider the round of error during the computation, we accept a threshold of \num{e-2} as the difference between the fault injected one and the fault-free one. As a result, we will define this case as SDC, if the ``min'' value of the last step is in [82.82, 82.83]; otherwise as detected. Figure~\ref{fig:mon_det} is a typical example of the case where the min is beyond such range. For the cases where the target file cannot be created, they are defined as crash.

For each benchmark application, we first answer the question: ``How do the HPC applications react to different SSD-related failures that propagate to the application data?". Then, we conduct a detailed analysis to reason about cases for the different failure categories. 

\subsection{HDF5 Metadata Fault Injection} 
We describe the approach to investigate how a storage-related error occurring in the HDF5 metadata affects the application's error resilience. 

This approach is based on the following observation:
when an HDF5 file is created, the HDF5 library first locks the file to prevent the concurrent writes from other processes, and then performs multiple writes to store the raw data; after that, it packs all metadata and write them to the file and unlocks the file for later access.

Based on this procedure, \ffis identifies the specific write operation for metadata (i.e., the penultimate \textit{fwrite}) and then perform a fault injection starting from the offset value specified by the \textit{fwrite} and till the end of the buffer byte-by-byte. Note that we refer to the HDF5 File Format Specification \cite{hdf5-byte} to capture the field information of each metadata byte and analyze the results accordingly.  


\section{Evaluation Results}
\label{sec:eval}
In this section, we present our fault injection results on the target HPC applications, evaluate the fault robustness of the applications, and discuss some observations and insights. 

\subsection{Results for Faults Affecting HDF5 Metadata}

In this section, we present the evaluation result of the cases where the fault affects the HDF5 file metadata for the Nyx cosmological simulation application. 

We first show the overview of the results of fault injection in the HDF5 metadata of Nyx dataset in Table~\ref{table:Nyx-1}. 
As shown in the table, 0.2\% of the total cases lead to SDC, which means that they cannot be detected by the post-analysis procedure; 85.7\% of the total cases have the same as the original (fault-free) data (i.e., benign); and 14.1\% of the total cases cause the Nyx application to crash (mainly due to the exceptions thrown by the HDF5 library, indicating the values in the fields become unjustified by the library).

Below, we analyze the benign and SDC cases and associate the results with the injected fields of the metadata  in detail. 

\textbf{\textit {Analysis of the benign cases:}} there are two types of fields contributing to the benign cases that are the dominant types of the HDF5 file format related failures:

\begin{enumerate}
    \item Based on the HDF5 format specification and the HDF5 library’s default space allocation policy, most of bytes of the total metadata belong to reserved fields, alignment space between fields, and space reserved for future metadata. For example, the B-tree nodes that accounts for 72\% of the total metadata space, can be partially full (i.e. 10\%), which leads to the situation that much of the space in the file metadata remains unused. The faults affecting these bytes would not make any impact on either the integrity of the HDF5 files or the post-analysis procedure.
    \item There are certain fields that exhibit resilient behaviors for the HDF5 files. Some examples are as follows:
    \begin{itemize}
        \item \textsc{Bit precision}: It represents the field of objHeader.dataType.floatingPointProperty and it shows the number of bits of precision of the floating-point value within the datatype.   
        \item \textsc{Bit offset}: This maps to the objHeader.dataType.floatingPointProperty field and represents the bit offset of the first significant bit of the floating-point value within the datatype. Therefore, the faults affecting this field might not cause much change to the floating-point value. 
        \item \textsc{Size}: The size field defines the objHeader.layout.contiguousStorageProperty. This field contains the size allocated to store the raw data, in bytes. We observe that if a fault modifies the size to a bigger value, the application would still produce the correct output, otherwise a crash would occur. 
    \end{itemize}
\end{enumerate}


\textbf{\textit {Analysis of the SDC-causing fields:}} It is also worth noting that there are 6 potential metadata fields that may cause SDC, including \textit{Bit-5 of Mantissa Normalization}, \textit{Exponent Location}, \textit{Mantissa Location}, \textit{Mantissa Size}, \textit{Exponent Bias}, and \textit{Address of Raw Data (ARD)}. We describe each possible SDC case based on their errors in terms of halo mass, halo locations, number of halos, and the average value of input data in post-analysis (as shown in Table~\ref{table:Nyx-pos}), and select 3 typical cases to visualize.

\begin{figure}[h]
     \centering
     \begin{subfigure}[t]{0.32\linewidth}
         \centering
         \includegraphics[width=\linewidth]{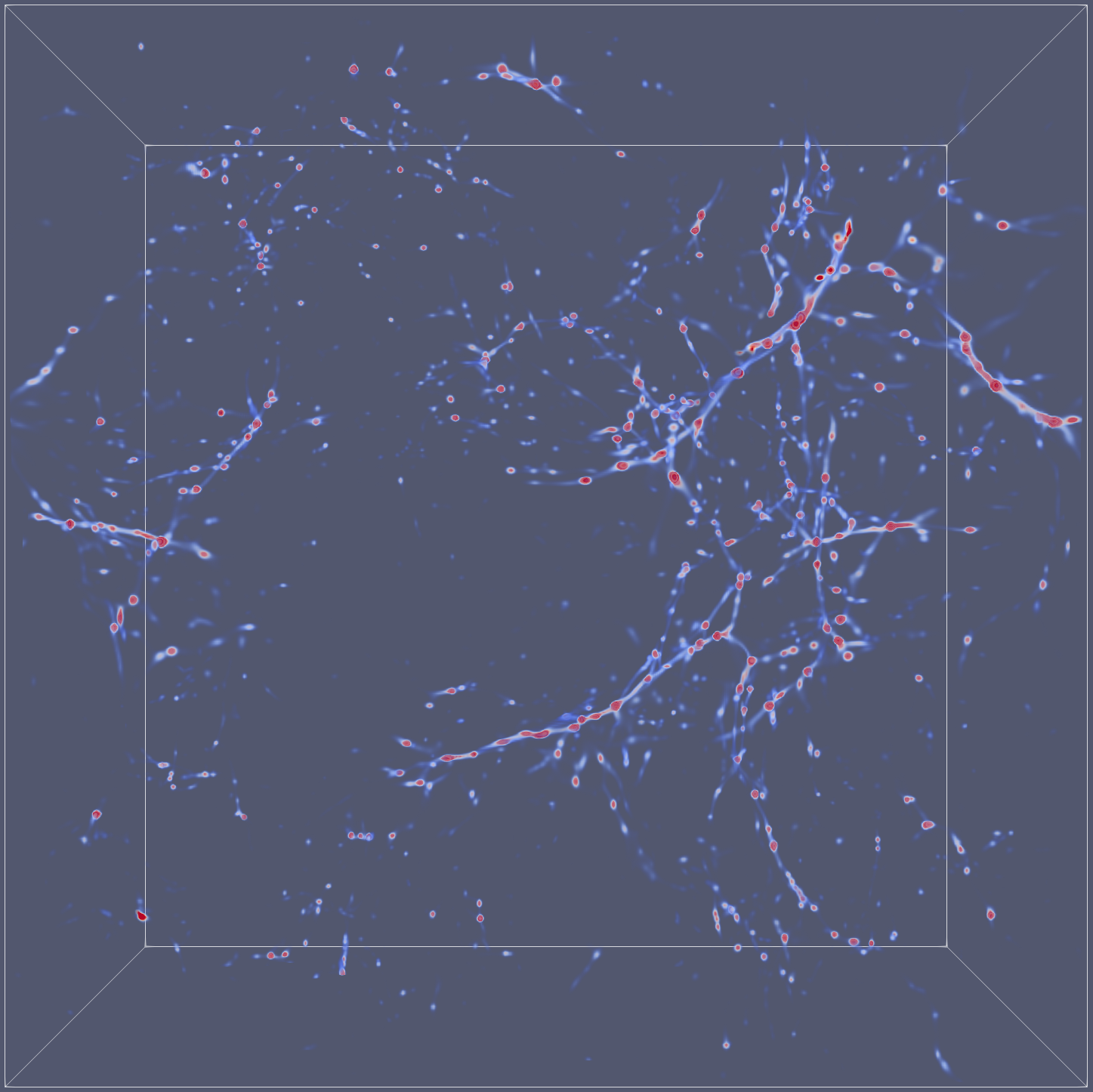}  
         \caption[t]{Original}
         \label{nyx:sub-1}
     \end{subfigure}
     \begin{subfigure}[t]{0.32\linewidth}
         \centering
         \includegraphics[width=\linewidth]{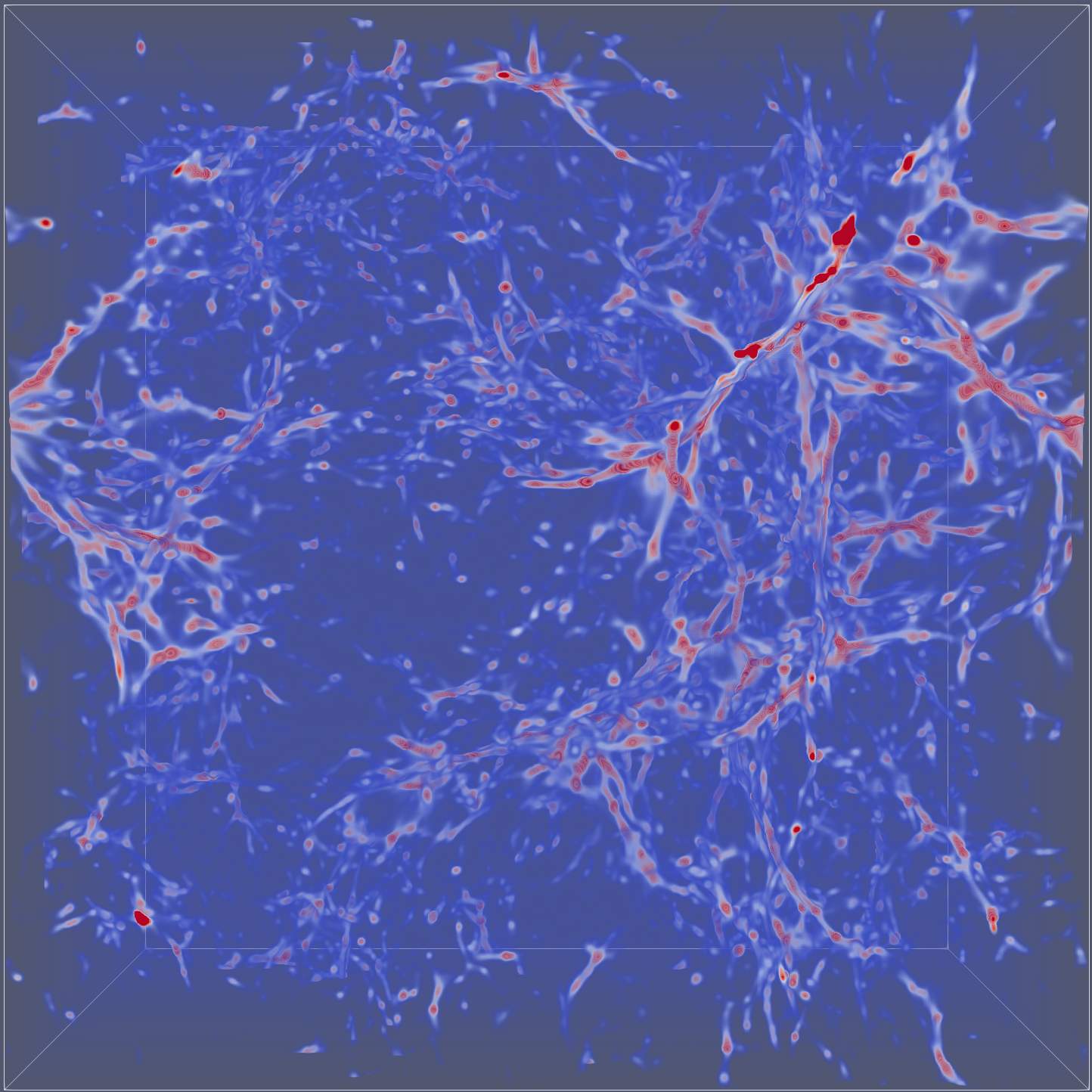}
         \caption{Exponent Bias}
         \label{nyx:sub-3}
     \end{subfigure}
          \begin{subfigure}[t]{0.32\linewidth}
         \centering
         \includegraphics[width=\linewidth]{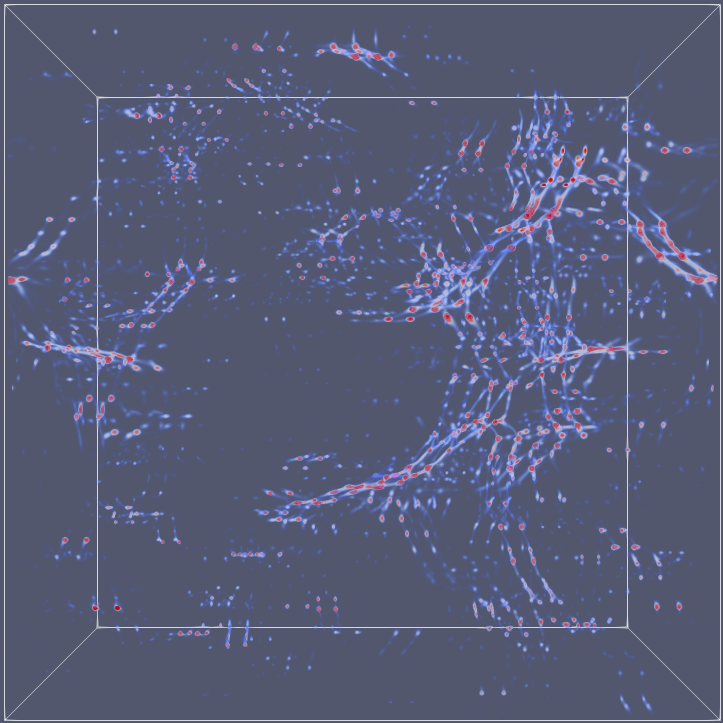}  
         \caption{ARD}
         \label{nyx:sub-2}
     \end{subfigure}
        \caption[t]{Visualization of typical SDC cases. A faulty Exponent Bias (b) scales up the input data; a faulty ARD (c) shifts the input data.}
        \vspace{-2mm}
        \label{fig:nyx4}
\end{figure}

When there is a faulty \textit{Exponent Bias}, as shown in Figure~\ref{nyx:sub-3}, the mass of all found halos change by the same multiple (i.e., scaled), while all locations are unchanged. 

When a Faulty \textit{ARD} occurred, the input data will be shifted, as shown in Figure~\ref{nyx:sub-2}. This causes all found halos' locations to shift, while the mass stays unchanged.

\begin{figure}[h]
     \centering
     \begin{subfigure}[t]{0.49\linewidth}
         \centering
         \includegraphics[width=\linewidth]{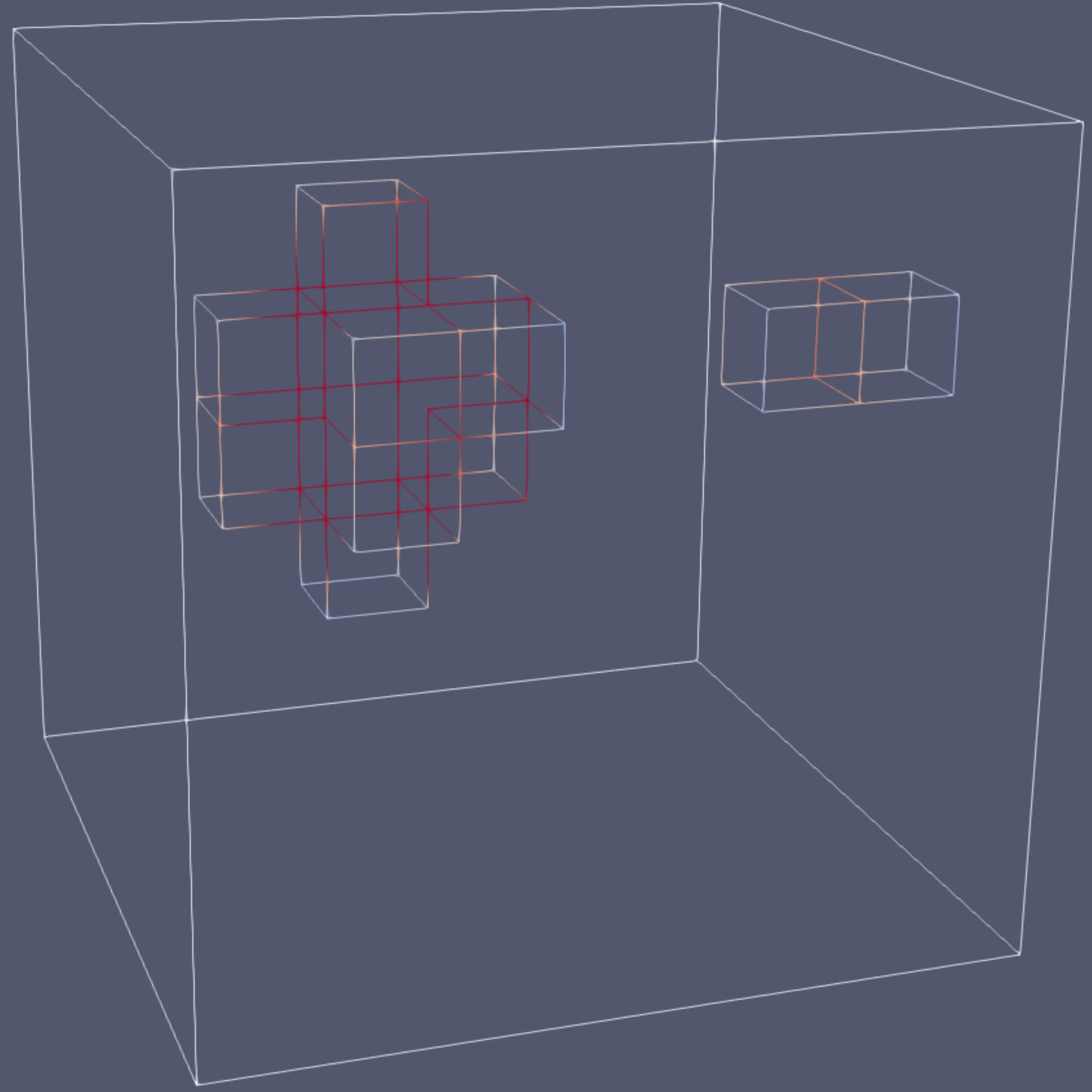}  
         \caption[t]{Original}
         \label{nyx:nicezoom}
     \end{subfigure}
     \begin{subfigure}[t]{0.49\linewidth}
         \centering
         \includegraphics[width=\linewidth]{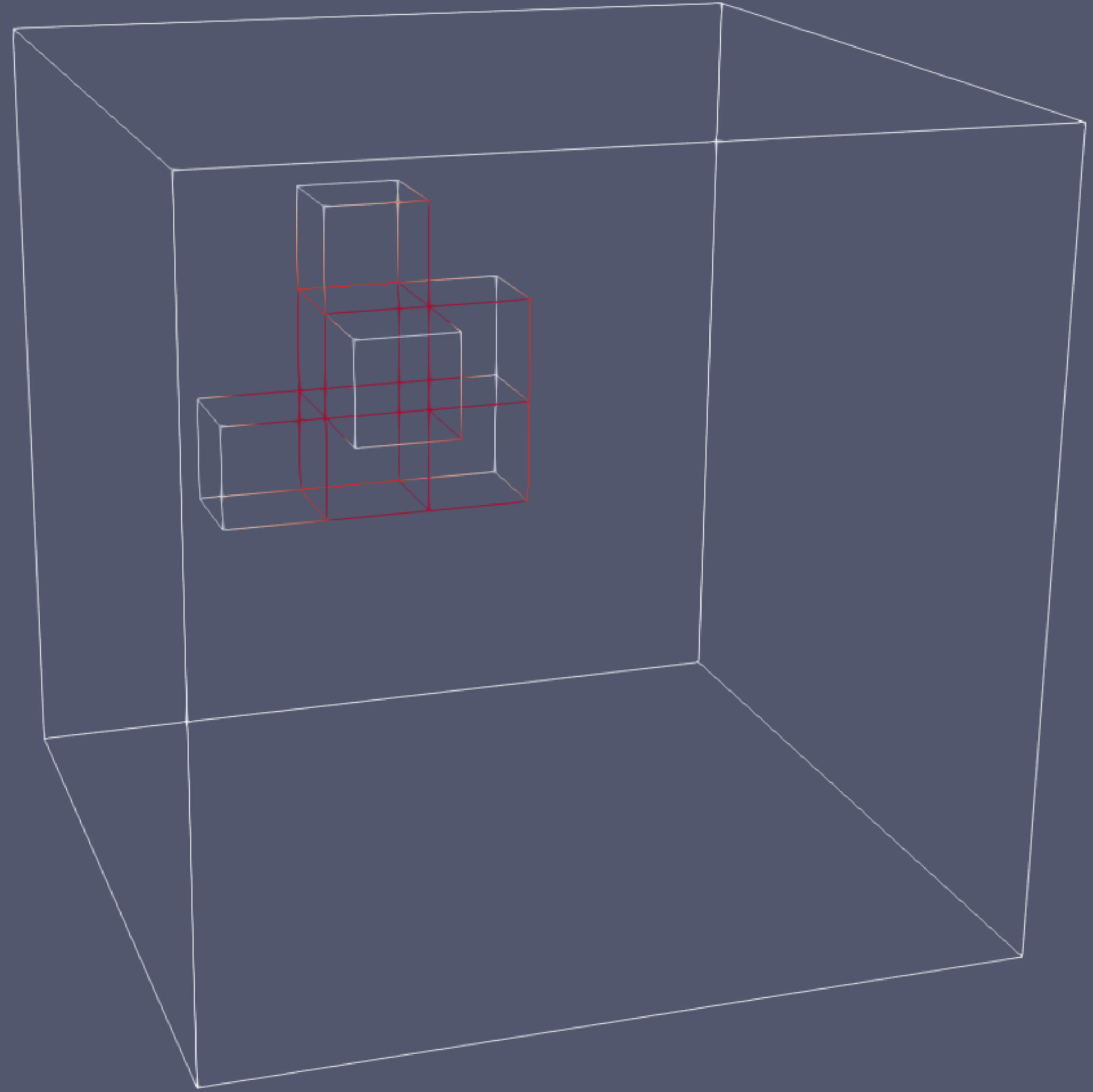}
         \caption{Faulty}
         \label{nyx:badzoom}
     \end{subfigure}
        \caption[t]{Visualization of a halo with a faulty Mantissa Size field (right). \small{A box indicates a halo cell candidate that meets the threshold. In the faulty case, the number of halo cell candidates is reduced compared to the original case (left) thus there are not enough halo candidates to form a halo.}}
        \label{fig:nyxzoom}
\end{figure}

When \textit{Exponent Location}, \textit{Mantissa Location}, or \textit{Mantissa Size} is changed, some halos will be missing or some additional halos will be found. Figure~\ref{fig:nyxzoom} shows two halos found by the golden run and SDC run caused by the fault in \textit{Mantissa Size} field.

Thus, we conduct an in-depth study on these metadata fields, and propose a novel and specific approach to potentially help HDF5 library detect and correct the faulty value residing in those fields. 
\begin{enumerate}
    \item \textit{Detection approach to identify which field is incorrect:} We first introduce the detection mechanism to identify which field may be affected by the fault. For Nyx, we observe that the average value of original input data in Nyx should remain 1 due to the law of mass conservation. Therefore, if the average value of the input data is not 1, which indicates that there might be an error in one of these fields. Next, we examine the actual average value, and classify the value to speculate where the fault resides:
    \begin{itemize}
        \item If the  average value of the input  data is the power  of 2, the \textit{Exponent Bias} field might be erroneous. 
        \item If the average value is between 1 and 2, the ``datatype massage'' in metadata can be used to determine whether there is a fault within \textit{Exponent Location}, \textit{Mantissa Location}, or \textit{Mantissa Size} fields.
    \end{itemize}
    
    \item \textit{Correction methodology:} 
    \begin{itemize}
        \item To correct the fault in \textit{Exponent Bias} field, one can re-scale the value of the \textit{Exponent Bias} field based on the average value observed. For example, our experiment shows that if the average value becomes 4096 after fault injection, the \textit{Exponent Bias} changes from 0x0000007f (the \textit{Exponent Bias} for IEEE single-precision floats) to 0x00000073, and this value can be corrected via added by 12(i.e. $2^{12}$)
        \item Due to the constraint of these three fields in floating-point representation (e.g., 8-bit exponent is saved next to 23-bit mantissa in single precision), users can fix the fault by making sure that (1) the value of \textit{Exponent Location} is equal to the value of \textit{Mantissa Size} (e.g., 23), and (2) the value of \textit{Mantissa Size} plus the value of \textit{Exponent Size} (e.g., 8) is equal to the precision number minus 1 (e.g., 31, due to 1 sign bit). 
    \end{itemize}

\end{enumerate}





However, for faulty \textit{ARD}, unlike the above SDC cases, users cannot determine whether there is a fault based on the average value of input data (because it remains 1), causing a severe impact on the post-analysis result.  Thus, we need to introduce a protection mechanism to prevent this harmful SDC case. We note that the metadata is saved followed by data in the HDF5 file format, the \textit{ARD} is exactly equal to the size of metadata. As a result, we can efficiently detect and correct the faulty  \textit{ARD} by changing it back to the metadata size.

\begin{table}
  \caption{Output classification of faulty metadata.}
  \label{table:Nyx-1}
  \centering
    \begin{tabular}{@{}lll@{}}
    \toprule
    \textbf{Fault Types} & \textbf{Case Number} & \textbf{Example Metadata Fields and Bytes}                                                                                                                                            \\ \midrule
    SDC                  & 4 (0.2\%)             & \begin{tabular}[c]{@{}l@{}}Bit-5 of Mantissa Normalization,\\ Exponent Location, Mantissa Location,\\ Mantissa Size, Exponent Bias,\\ Address of Raw Data (ARD)\end{tabular}  \\ \midrule

     Benign                 & 2085 (85.7\%)             & \begin{tabular}[c]{@{}l@{}}Bit Offset, Bit Precision, Size\\ Reserved and unused bytes, \\ other trifling fields, etc. \end{tabular}  \\ \midrule
    
    
    Crash                & 343 (14.1\%)          & \begin{tabular}[c]{@{}l@{}}Version \# of Data Object Header,\\ Version \# of Data Object Header Message,\\ Symbol Table Node signature,\\ B-tree signature, etc.\end{tabular} \\ \bottomrule
    \end{tabular}
\end{table}

\begin{table*}[h]
\caption{Description of erroneous post-analysis result in Nyx with faulty metadata fields causing SDC.}
\label{table:Nyx-pos}
\begin{tabular}{@{}lllllll@{}}
\toprule
\textbf{Metrics} & \textbf{Mantissa Normalization}                                                      & \textbf{Exponent Location}                                                    & \multicolumn{2}{l}{\textbf{Mantissa Location / Mantissa Size}}                                                                     & \textbf{Exponent Bias}                                                 & \textbf{ARD}                                                                 \\ \midrule
Halo Mass        & \begin{tabular}[c]{@{}l@{}}Mass of all halos\\ have changed\end{tabular}             & \begin{tabular}[c]{@{}l@{}}Mass of all\\ halos have changed\end{tabular}      & \multicolumn{2}{l}{Mass of all halos has changed}                                                                                  & \begin{tabular}[c]{@{}l@{}}Mass of all halos\\ was scaled\end{tabular} & Unchanged                                                                    \\ \midrule
Halo Location    & \begin{tabular}[c]{@{}l@{}}45\% of halos have\\ changed their locations\end{tabular} & \begin{tabular}[c]{@{}l@{}}Locations of all\\ halos have changed\end{tabular} & \multicolumn{2}{l}{Locations of most halos have changed}                                                                           & Unchanged                                                              & \begin{tabular}[c]{@{}l@{}}Locations of all\\ halos are shifted\end{tabular} \\ \midrule
Halo Number      & Increased by 24\%                                                                    & Increased by 20\%                                                             & \multicolumn{2}{l}{\begin{tabular}[c]{@{}l@{}}Changed (ranging from 320 to 527\\ based on our massive experiments)\end{tabular}}   & Unchanged                                                              & Unchanged                                                                    \\ \midrule
Average Value    & Changed to 0.55                                                                      & Changed to 1.04                                                               & \multicolumn{2}{l}{\begin{tabular}[c]{@{}l@{}}Changed (ranging from 1.04 to 1.55\\ based on our massive experiments)\end{tabular}} & \begin{tabular}[c]{@{}l@{}}Scaled by a power\\ of two\end{tabular}     & Unchanged                                                                    \\ \bottomrule
\end{tabular}
\vspace{-4mm}
\end{table*}




\textbf{Insight on HDF5 Metadata}
We observe that due to the high fault tolerance of Nyx's post-analysis, the faults in HDF5 metadata will have a relatively small impact on the halo-finder analysis result in general. But there is still a low probability of SDC case that will result in a very serious impact. Therefore, we propose an average-value-based method for users to detect SDC faults in Nyx's post-analysis and protect the key fields in HDF5 metadata, which further improves the fault tolerance of Nyx application against storage faults.

We also note that the baryon density field in Nyx can be easily compressed (i.e., compression ratio ranging from tens to hundreds) \cite{jin2020understanding, jin2021adaptive}, thus the importance of metadata would be greatly raised due to its increasing portion in the whole file. And since some metadata fields are related to each other, certain faults in the metadata can be detected and corrected as aforementioned; in other words, as the metadata of HDF5 file format itself has a certain degree of redundancy (correlation), we do not choose to replicate the metadata.



\subsection{Results for Faults Affecting Application Data}
In this section, we evaluate error resilience of Nyx, QMCPACK, Montage when the errors affect the application data. 
\begin{figure}
    \centering
    \includegraphics[width=0.5\textwidth]{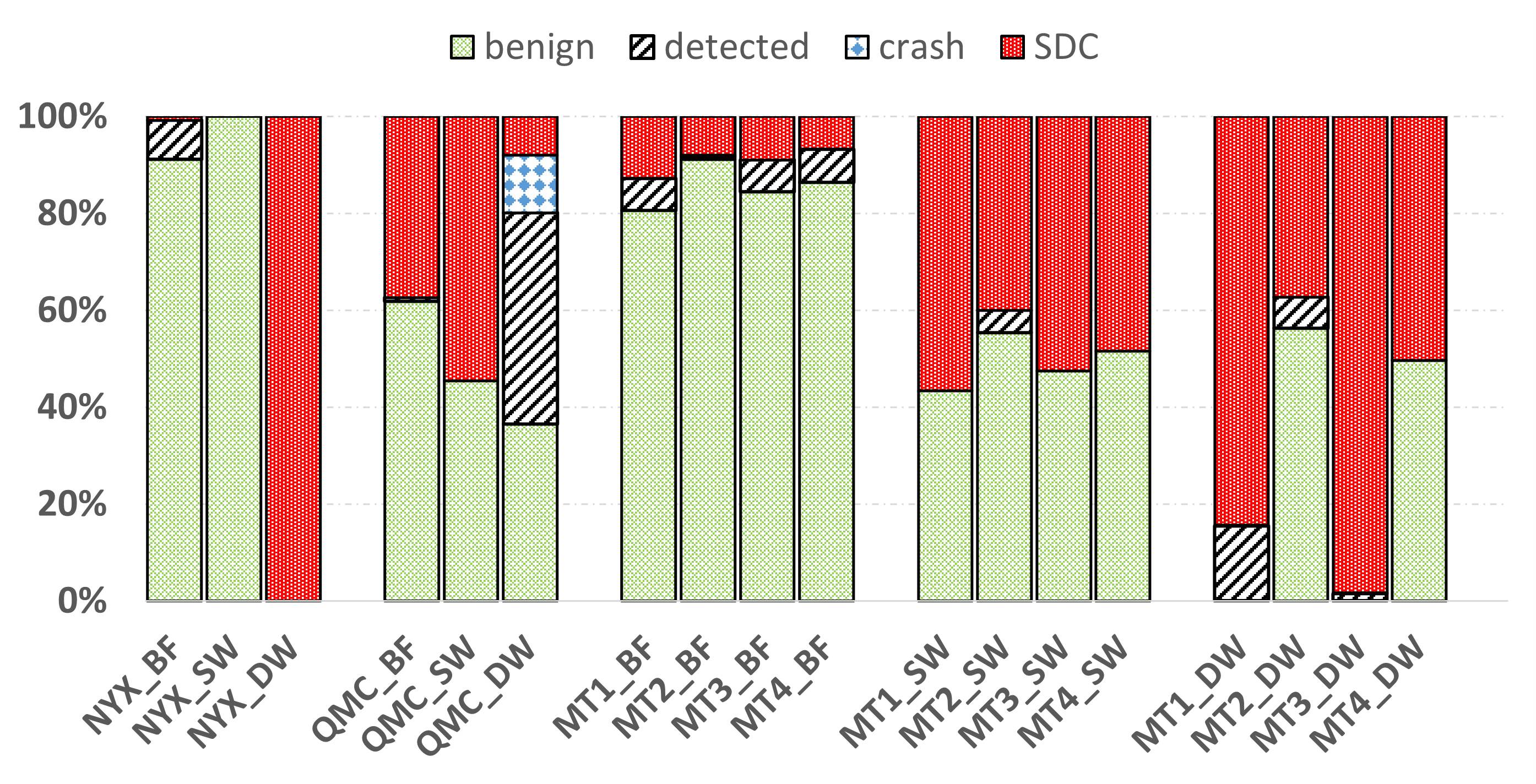}
    \vspace{-6mm}
    \caption{Characterization result of I/O faults with Nyx, QMCPACK, and Montage. \small{Note that all SDC cases with Nyx will be changed to detected cases after using the average-value-based method. ``NYX'', ``QMC'', and ``MT1/2/3/4'' represent Nyx, QMCPACK, and different stages in Montage, respectively. ``BF'', ``SW'', and ``DW'' represent \textsc{bit flip}, \textsc{shorn write}, and \textsc{dropped write}, respectively.}}
    \label{fig:merge}
\end{figure}

\paragraph{Nyx}

Figure~\ref{fig:merge} shows the result of halo-finder analysis on Nyx's baryon density variable with injected faults. For \textsc{bit flip}, there are 91.1\% cases that Nyx produces the exact results compared to the golden, and only 0.8\% SDC cases occurred, which is the lowest SDC rate among the three applications. For \textsc{shorn write}, surprisingly, there is no affect on the halo-finder analysis (i.e., all the cases are Benigns). And for \textsc{dropped write}, 1000 out of 1000 injected faults cause SDC outcomes.

We explain the reasons for the differences in the impact of different fault types on the post-analysis results. As aforementioned, the halo-finder algorithm searches for the halos from all the simulated data, with the following two criteria: (1) the mass of an object(s) must be greater than a threshold (e.g., 81.66 times the average mass of the whole dataset) to become a halo cell candidate \cite{jin2020understanding, jin2021adaptive}, and (2) there must be enough halo cell candidates in a certain area to form a halo. Below, for each fault type, we explain in details how each fault type potentially affects the halo-finder procedure.

\begin{figure}
    \centering
    \includegraphics[width=0.45\textwidth]{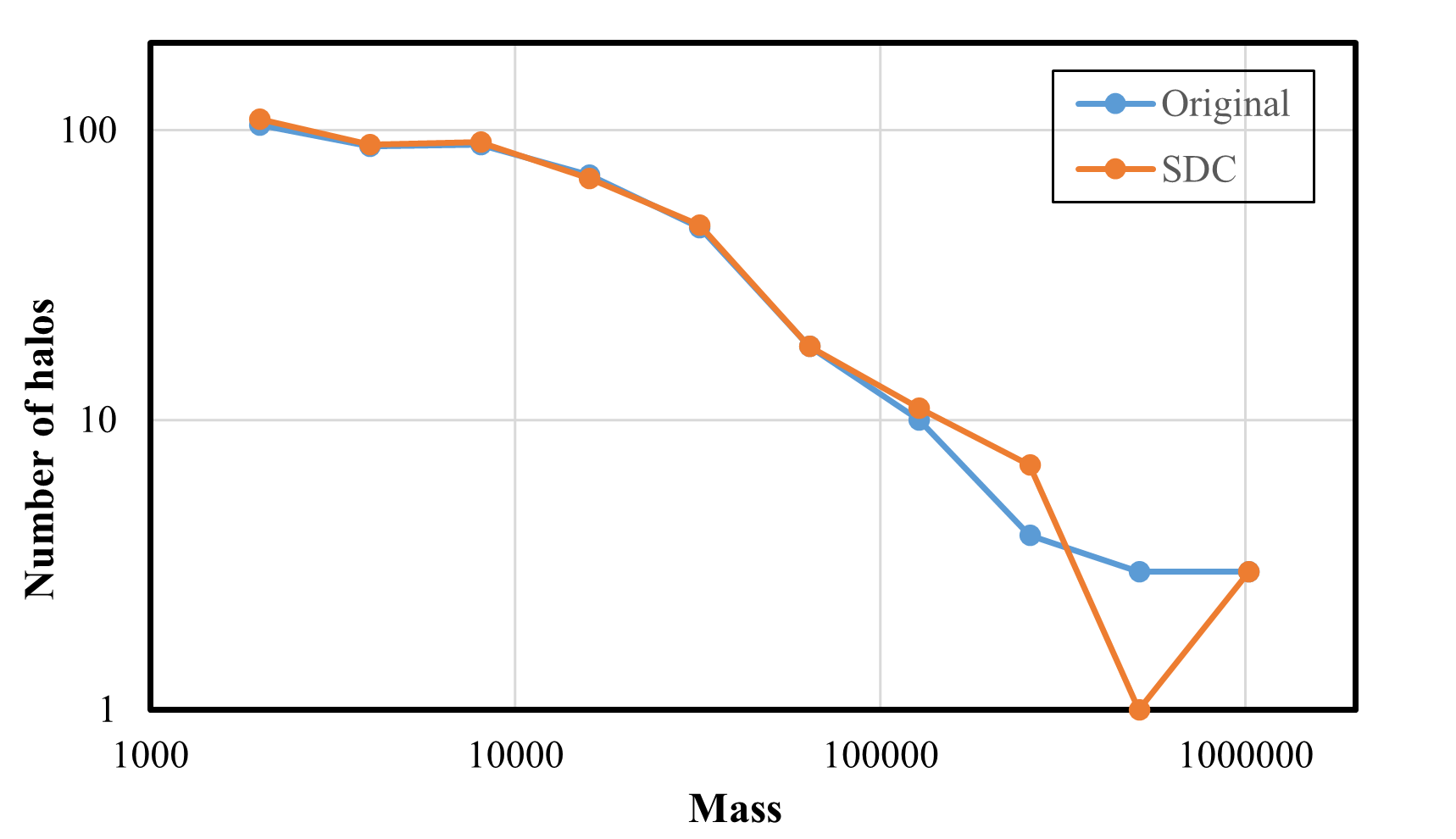}
    \vspace{-2mm}
    \caption{Comparison of halo-finder analysis on original and faulty baryon density data in Nyx. \small{The x-axis represents the halo mass. the left y-axis represents the counts of halos.}}
    \label{fig:nyxcurve}
\end{figure}

\begin{enumerate}
    \item \textbf{Bit Flip} Although the bit-flip error only affects one data point of Nyx, the halo finder may fail to find all the Halos. This is mainly because when the mass of a certain point in the dataset changes sharply, the average value of the input dataset changes accordingly, which causes the mass of all points in the dataset to be less than the threshold. The consequence is that no halo candidates can be found (i.e. detected). On the other hand, when the change to the average value is not significant, the original halo candidates would still satisfy the searching criteria, but a particular point affected by the fault may cause the outcome to be slightly different than the golden run (i.e. SDC). Since most of the points do not participate in halos, the chance for the halo finder to produce the correct halos dominates the outcomes (i.e. benign). 
    \item \textbf{Shorn Write}  \textsc{Shorn write} replaces the unsuccessfully written data with the data that is within an order of magnitude difference from the original data, which results in all faults being very small. We confirmed this by checking the values of the changed file. Thus, due to the characteristics of halo finder, these ``moderate'' faults are all mitigated.
    \item \textbf{Dropped Write} 
    Unlike \texttt{shorn write}, a \textsc{dropped write} fault drops a large piece of data, leading to a change in the average mass of the dataset. Thus the halo-finder algorithm would find a different number of cells in a halo compared to the golden run, and this can not be mitigated by the halo finder procedure. Figure \ref{fig:nyxcurve} shows the distribution of the mass for the identified halos on faulty and original data. Note that the SDC curve is different from the original curve, especially when the mass is relatively large, because halos with larger mass have more halo cells and are more susceptible to \textsc{dropped write}.
\end{enumerate}

\textbf{Observation and Insight} Although \textsc{dropped write} has a 100\% of SDC rate, all the SDC cases in our experiment can be detected by using the average value, because the average value is reduced by at least 0.1\% (e.g., less than 0.9983) for all the SDC cases. Thus, we recommend Nyx users to keep using the average-value-based method to protect the data from storage faults with respect to halo-finder analysis. Moreover, as explained above, an important reason is that the halo-finder process is sensitive to large deviations, while the small deviation introduced by \textsc{bit flip} would not compromise the outcomes. 
As a result, Nyx is highly resilient to storage faults after using the average-value-based method. 

\paragraph{QMCPACK}


Figure~\ref{fig:merge} illustrates the result of fault injection with QMCPACK. Through the figure, we can observe that the fault injection results with different types of faults are similar. Specifically, in each fault model, there are about 50\% of the faults to be SDC; in \textsc{bit flip}, it even reaches 60\%. Also, in \textsc{bit flip} 37\% of the faults are SDC and 0.8\% of the faults are detected. In \textsc{shorn write} 54\% of the faults are SDC with no case is detected. In \textsc{dropped write} 8\% of the faults are SDC, 43\% of the faults are detected and 12\% of the faults are crash.
This is because floating-point numbers can mitigate fault to a certain extent.



Compared to \textsc{bit flip}, \textsc{shorn write}.  has a higher percentage of SDC and hence a higher impact on QMCPACK, which is reasonable. This is because unlike \textsc{bit flip} only changes two bits, \textsc{shorn write} could affect 512 Bytes. 
It is worth noting that all the \textsc{shorn write} faults are SDC while some of \textsc{bit flip} faults are detected. This is because, on one hand, \textsc{bit flip} simply flip two bits in a floating-point number; if the fault happens to occur in the most significant bits, then \textsc{bit flip} will greatly affect the floating-point value, thus greatly affecting the final result.
On the other hand, the principle of \textsc{shorn write} is to discard the portion of data and replace them with random values. 
From the result, the difference between the replacement value and the original value is minimal.



Similar to \textsc{shorn write}, the impact of \textsc{dropped write} is milder than that of \textsc{bit flip}. But compared to \textsc{shorn write}, the number of bits changed by \textsc{dropped write} is larger.
This results in that the final output of \textsc{dropped write} deviates more from the correct energy compared to \textsc{shorn write}. 
Although the difference is not very large, 43.6\% of the cases exceed the threshold (i.e., 10\%) that we set for SDC. Therefore, the detected cases in \textsc{dropped write} are more than the other two types of faults.



\textbf{Observation and Insight} QMCPACK is not resilient to \textsc{bit flip} and \textsc{shorn write}, as they have a high likelihood to cause a minor deviation in the outcomes. That said, to improve the error resilience of QMCPACK, more advanced techniques guided by more domain knowledge need to be considered. 


\paragraph{Montage}
We report the fault injection results of Montage in Figure~\ref{fig:merge}.
We select the most four I/O-intensive stages for our fault injection: (1) mProjExec for reprojecting each image, (2) mDiffExec for subtracting each pair of overlapping images and creating difference images, 
(3) mBgExec for applying background matching to each reprojected image, (4) mAdd for generating a mosaic from reprojected images.



We investigate the characteristics of potential fault propagation in Montage, that is, to study if the impact of the faults has time dependency.
We execute BF1 (mProjExec), BF2 (mDiffExec), BF3 (mBgExec), and finally execute BF4 (mAdd) sequentially, and the results of fault injection experiments for different are presented in Figure~\ref{fig:merge}. By observing the number of SDC and benign cases in each Montage stage for \textsc{bit flip} and \textsc{shorn write}, we find that in the \textsc{bit flip} cases all the numbers stay relatively stable across different stages (e.g., 12.8\%, 8\%, 9\%, 6.8\%). In the \textsc{shorn write} cases, the fluctuation of the SDC and benign cases vary slightly across different stages (e.g., 56.6\%, 40\%, 52.5\%, 48.5\%), while there is no indication on a statistically significant trend observed. As of \textsc{dropped write}, the SDC and benign cases vary in the same way as \textsc{shorn write}, but more drastically (e.g., 83.5\%, 37.3\%, 98.3\%, 50.4\%). 

Interestingly, in stage two (i.e., mDiffExec), the data (diffdir) generated is not directly applied to the subsequent steps, but to be used to calculate plane-fitting coefficients for each difference image through the second stage, which potentially be mitigated in the process of extracting coefficients. This explains why the SDC rate in the second step is the lowest among the three types of faults.

\begin{figure}[h]
     \centering
     \begin{subfigure}[b]{0.49\linewidth}
         \centering
         \includegraphics[width=\linewidth]{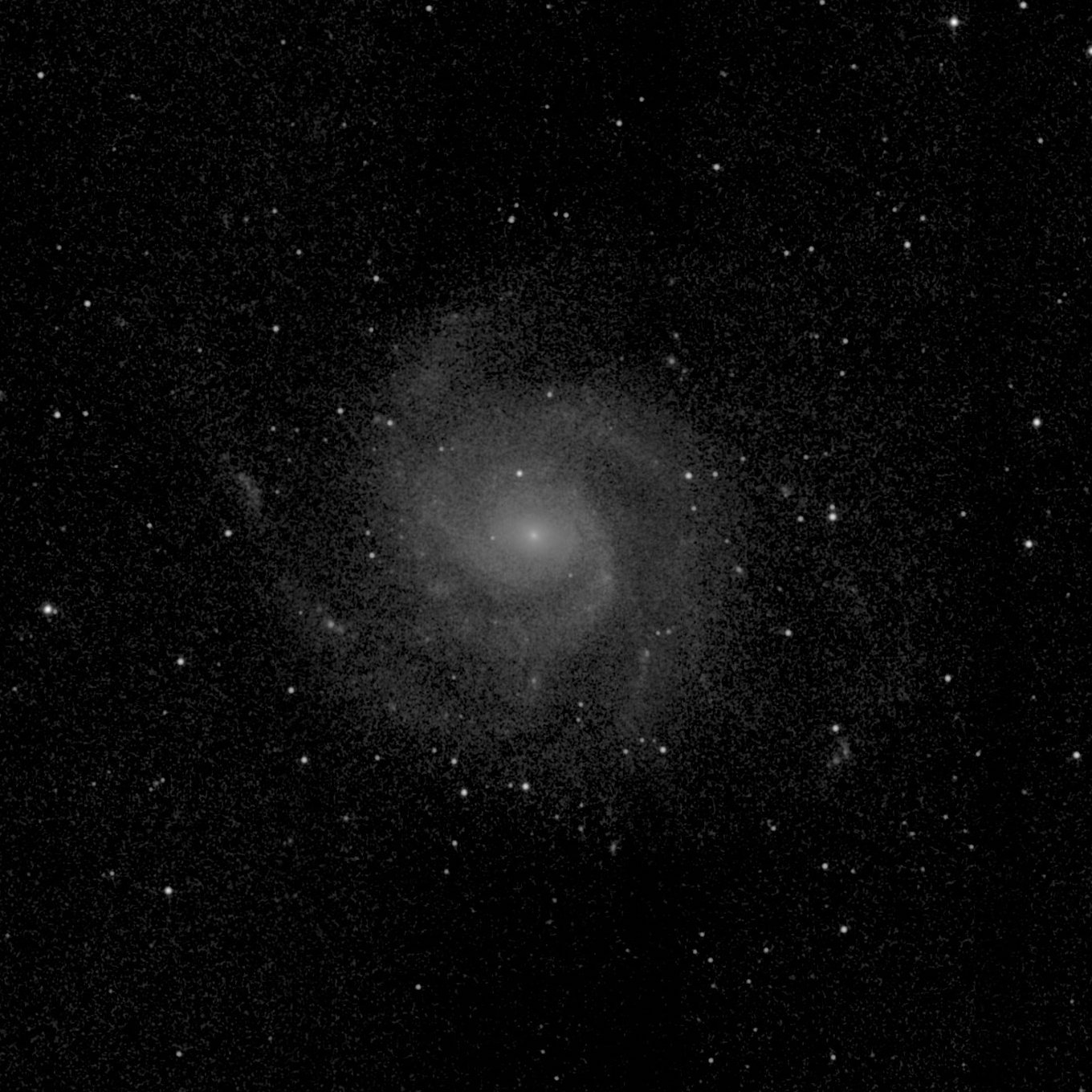}
         \caption{Original}
         \label{fig:monori}
     \end{subfigure}
     \begin{subfigure}[b]{0.49\linewidth}
         \centering
         \includegraphics[width=\linewidth]{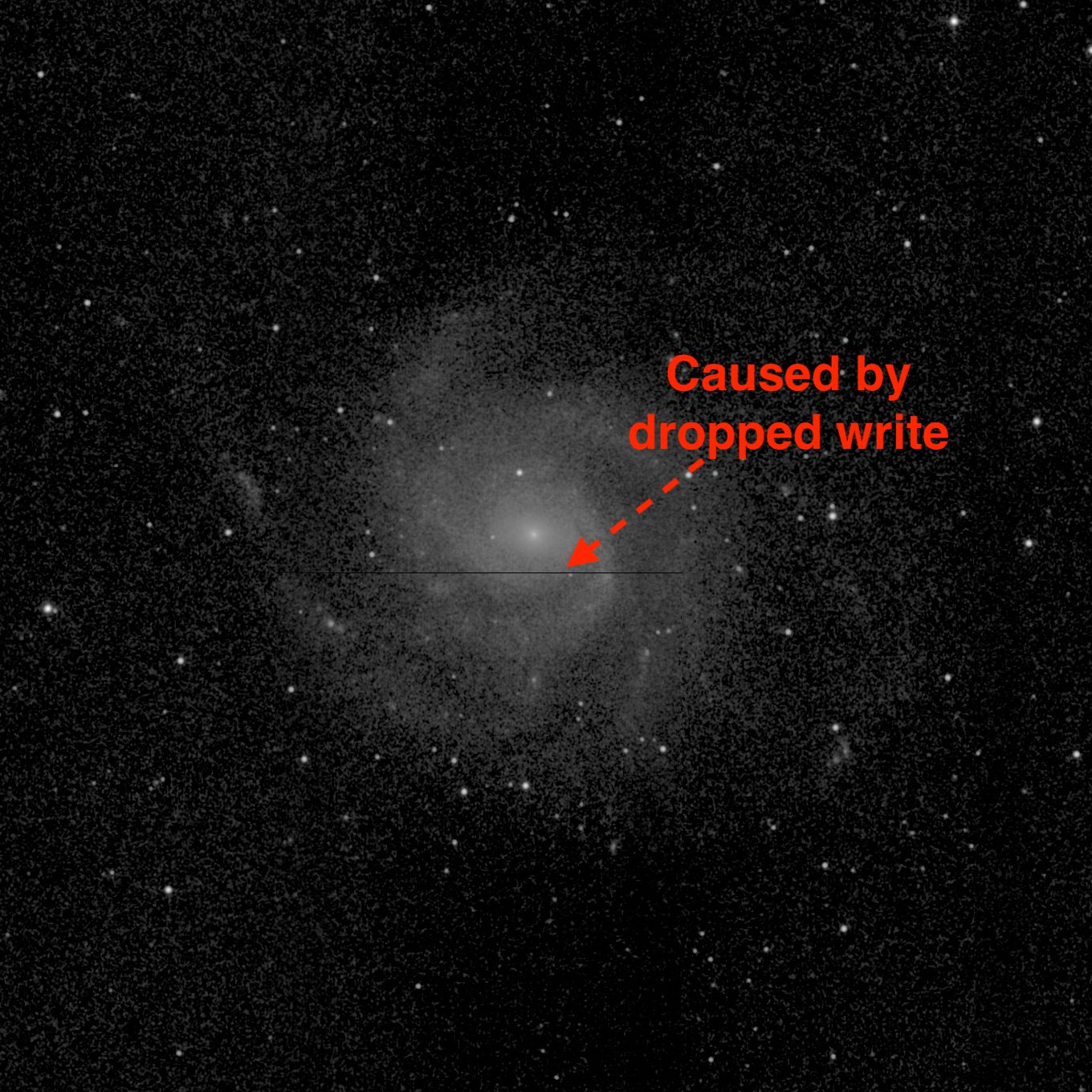}
         \caption{Faulty (detected)}
         \label{fig:mon_det}
     \end{subfigure}
        \caption{A typical faulty image due to \textsc{dropped write}.}
        \label{fig:montage_example}
\end{figure}


Figure \ref{fig:montage_example} shows a typical example of faculty image due to \textsc{dropped write}. It can be seen that there is a black line in the middle of the vortex, which is caused by missing a large piece of data due to \textsc{dropped write}.



\textbf{Observation and Insight} For Montage we observe the relative small fluctuation on the SDC rates for \textsc{bit flip} and \textsc{shorn write} across the four stages. This indicates that instead of ``fault masking" or ``fault propagation" behaviors, different Montage stages seem to bound the faults and the error resilience on each stage decouples from each other.



\section{Related work}
\label{sec:related}

\textbf{Characterization of HPC Application Error Resilience } There have been a large body of studies focusing on estimating the error resilience of HPC applications via fault injection experiments. For example, Ashraf et al.~\cite{Ashraf2015} propose a fault injection and propagation framework that tracks the transient hardware faults within a process and across MPI processes. Calhoun et al. used FlipIt~\cite{flipit}, an LLVM-based FI tool, to investigate how corrupted data propagate through a specific HPC computation kernel~\cite{spmv}. There is also another line of work~\cite{gpuqin, sassfi,nvbitfi,llfigpu} that investigates the error resilience of GPGPU applications targeting high-performance accelerators. However, non of these tools are capable of modeling SSD-related failures with software-implemented faults. 

\textbf{SSD Failure Simulation} Xu et al.~\cite{ssd-bfrate} model the raw bit error rate at the disk level to quantitatively analyze unique error behaviors in SSDs. It investigates data-level error tolerance for various applications via introducing different error rates at the SSD simulator. It found that for some applications under certain error rates, it is possible to disable ECC for those applications with acceptable level of error tolerance. Although their aim aligns similar to ours, the two studies differ as follows: \textit{(i)} their method is not practical for HPC applications due to the performance of the SSD simulator, which is the core implementation interface of their work; \textit{(ii)} they need further interpretation to understand the application's reliability on HPC systems that experience a collective error rate; \textit{(iii)} they did not consider other SSD-related failure types. 

\textbf{SSD Failures Affecting File Systems} PFault~\cite{pfault} is a general framework for analyzing the failure handling of parallel file systems. PFault automatically captures I/O commands on all storage nodes of lustre file system, issues realistic failure states and examine if lustre can detect and recover from such failure states. They randomly trash data on the local file system to emulate the failure state, which differs from our dedicated failure modeling techniques. In addition, Jeffer et al.~\cite{eval-ssd} conduct an extensive study on characterizing the resilience of various file systems (namely  \textit{ext4}, \textit{F2FS} and \textit{btrfs}) running on flash-based SSDs under SSD failures and evaluating the effectiveness of the recovery mechanisms of file systems. They are able to inject many types of faults as the manifestation of partial SSD failures, and present a detailed analysis over the file system error resilience characteristics. However, both~\cite{pfault, eval-ssd} do not focus on the application implication of the SSD-related errors, which is the main goal of this paper.

\textbf{Application-level Fault Injection for Storage} Ganesan et al.~\cite{errfs} build a fault injection framework called CORDS to study how the modern storage systems such as Redis, ZooKeeper, etc. handle the local fail system faults. Although CORDS also leverages the FUSE interface to plant the file system faults, there are fundamental differences between the two studies: \textit{(i)} CORDS models the faults without considering the source of the faults, which resulting significantly distinct fault models and fault injection methodology. For example, CORDS focuses on two types of faults: corruption and file system I/O errors. In the former case, they randomly modify the content of a read buffer, while in \ffis we delicately design the faults and injection methodology; for the latter case, applications receive signals to fail immediately, hence causing no data corruption; \textit{(ii)} we conduct a comprehensive study to understand the impact of errors on the scientific file format, and correlate such impact with the application behaviors.

\section{Conclusion}
\label{sec:conclusion}

This study focuses on the HPC applications affected by the SSD-related failures, and proposes the fault injection framework \ffis to study the impact of those failures on the applications. We conduct comprehensive fault injection experiments on three representative HPC applications from different domains, and our findings show that different applications exhibit dramatically error resilience behaviors against the same type of SSD-related faults (nearly no SDCs to more than 50\% of SDCs), while inside the same application (i.e. Montage), different stages of the application may have a similar error resilience characteristics against the same type of fault. 
Moreover, we unveil application-specifc behaviors operating on the most widely used scientific file format HDF5 and show the fault tolerance behaviors of the HDF5 library against errors affecting the HDF5 metadata. Finally, we propose a detection approach to identify which metadata field is potentially incorrect and corresponding correction methodology. 

\section*{Acknowledgement}
We would like to thank the anonymous reviewers for their feedback. This project was supported by DOE, Office for Advanced
Scientific Computing (ASCR) under
Concrete Ingredients for Flexible Programming Abstractions
Primary Project. The Pacific Northwest
National Laboratory is operated by Battelle for the U.S. Department of Energy under contract DE-AC05-76RL01830.
This work was also supported by the National Science Foundation under Grant OAC-2042084.
\clearpage
\bibliographystyle{IEEEtran}
\bibliography{citation}


\end{document}